\documentclass[aps,prb,twocolumn,superscriptaddress,floatfix]{revtex4-2}
\usepackage{graphicx}
\usepackage{color}
\usepackage{epsfig}
\usepackage{amsmath}
\usepackage{amssymb}
\usepackage{mathrsfs}
\usepackage{bm}
\usepackage{subfigure}
\usepackage{float}
\usepackage{url}
\usepackage{enumerate}
\usepackage{braket}
\usepackage{comment}
\usepackage{multirow}
\usepackage{extarrows}
\usepackage{xcolor}  
\usepackage[colorlinks,linkcolor=blue,anchorcolor=blue,citecolor=blue,urlcolor=blue]{hyperref}

\begin{document}
	\bibliographystyle{apsrev4-1}
	\renewcommand\arraystretch{1.5}
	\title{Variational corner transfer matrix renormalization group method for classical statistical models}
	
	\author{X. F. Liu}
	\affiliation{Department of Physics, Renmin University of China, Beijing 100872, P. R. China}
	\author{Y. F. Fu}
	\affiliation{Department of Physics, Renmin University of China, Beijing 100872, P. R. China}
	\author{W. Q. Yu}
	\affiliation{Department of Physics, Renmin University of China, Beijing 100872, P. R. China}
	\author{J. F. Yu}\email{yujifeng@hnu.edu.cn}
	\affiliation{Hunan University, Changsha 410082, P. R. China}
	\author{Z. Y. Xie}\email{qingtaoxie@ruc.edu.cn}
	\affiliation{Department of Physics, Renmin University of China, Beijing 100872, P. R. China}

	\date{\today}

	\begin{abstract}
		
		In the context of tensor network states, we for the first time reformulate the corner transfer matrix renormalization group (CTMRG) method into a variational bilevel optimization algorithm. The solution of the optimization problem corresponds to the fixed-point environment pursued in the conventional CTMRG method, from which the partition function of a classical statistical model, represented by an infinite tensor network, can be {efficiently evaluated}. {The validity of this variational idea is demonstrated by the high-precision calculation of the residual entropy of the dimer model, and is further verified by investigating several typical phase transitions in classical spin models, where the obtained critical points and critical exponents all agree with the best known results in literature. Its extension to three-dimensional tensor networks or quantum lattice models is straightforward, as also discussed briefly}.
		
	\end{abstract}
	\maketitle

	\section{Introduction}
	
	The study of the phase transitions among different phases has been always one of  the main focuses of condensed matter and statistical physics. For long time, people have believed that Landau's symmetry-breaking theory is accountable for all continuous phase transitions, and adequate to describe related different phases, until the Kosterlitz-Thouless (KT) transition\cite{kosterlitz1973jpc,kosterlitz1974jpc} was discovered in the two-dimensional $XY$ model in early 1970s, which is driven by topological excitations, or more specifically vortex-anti-vortex pairing instead of any symmetry breaking.
	{Unfortunately}, most many-body problems/models are too difficult or complicated to handle theoretically. Thus, a variety of numerical methods emerged, among which are two conventional candidates, Monte Carlo\cite{Metropolis1949jasta} and density matrix renormalization group (DMRG)\cite{White1992prl,White1993prb}. Around two decades ago, the tensor network state (TNS)\cite{Verstraete2004arxiv,ZhaoHH2010prb,Orus2020nrp} method developed rapidly as a generalization of DMRG for higher space dimensional problems, and {gradually became} one of the most powerful numerical tools for the phase transition study in both the classical statistical and quantum many-body systems.\cite{LiaoHJ2017prl,MeiJW2017prb,WangL2016prb,LeBlanc2015prx,Corboz2014prl,XieZY2012prb,YuJF2014pre,WangC2014prb}
	
	In {general}, the partition function of a classical statistical model, and the ground state wave function of a quantum lattice model, can be represented faithfully by an infinite tensor network, provided the interactions are of finite range. In such way, the physical problem is transformed to find an efficient method to contract the tensor network. There are several methods developed for this end\cite{Ran2020book}, such as {coarse-graining} tensor renormalization group\cite{Levin2007prl,XieZY2009prb,XieZY2012prb} {and its variational variants based on entanglement filtering\cite{GuZC2009prb,Evenbly2015prl,YangS2017prl,Bal2017prl}, boundary matrix product state (MPS)\cite{Vidal2003prl,Orus2008prb,Zauner2018prb}}, and corner transfer matrix renormalization group (CTMRG)\cite{Nishino1996jpsj,Corboz2014prl, Orus2009prb,Fishman2018prb} originating from Baxter's corner transfer matrix\cite{Baxter1982book} idea, to name a few.	Basically, these methods use iteration procedure to contract an infinite tensor network, and the physical quantities are dependent on the number of the kept states during the iteration process. While, generally this dependency is not monotonic, and this makes it difficult to guarantee the convergence of the investigated quantities.\cite{YangS2017prl,Corboz2012prx,XieZY2017prb}
	
	In this work, we reformulate the CTMRG method into a variational optimization algorithm (abbreviated as vCTMRG afterwards), see details in Sect.\ \ref{SectMethod} below.
	Using this new tool, we first study the ground state degeneracy of the dimer covering on the square lattice. Its residual entropy is calculated {with high-precision}, which {coincides excellently} with the theoretical exact value.
	We also investigate different types of phase transitions in classical statistical spin systems, as demonstrated by Ising, $q$-state Potts, and clock models on the square lattice. The obtained critical points and critical exponents all agree well with the predictions in literature, and the effectiveness and efficiency of this method are verified.
	
	\section{Method}\label{SectMethod}
	
	We have followed the idea of corner transfer matrix of Baxter\cite{Baxter1982book}, but in context of tensor networks\cite{Jahromi2018prb}. As mentioned above, the partition function of a classical statistical model on the square lattice can be represented as an infinite TNS, composed of identical local tensors $T$. {For convenience, the model is assumed isotropic, as in most classical spin models}.
	\begin{figure}[htbp]
		\centering
		\includegraphics[width=0.48\textwidth]{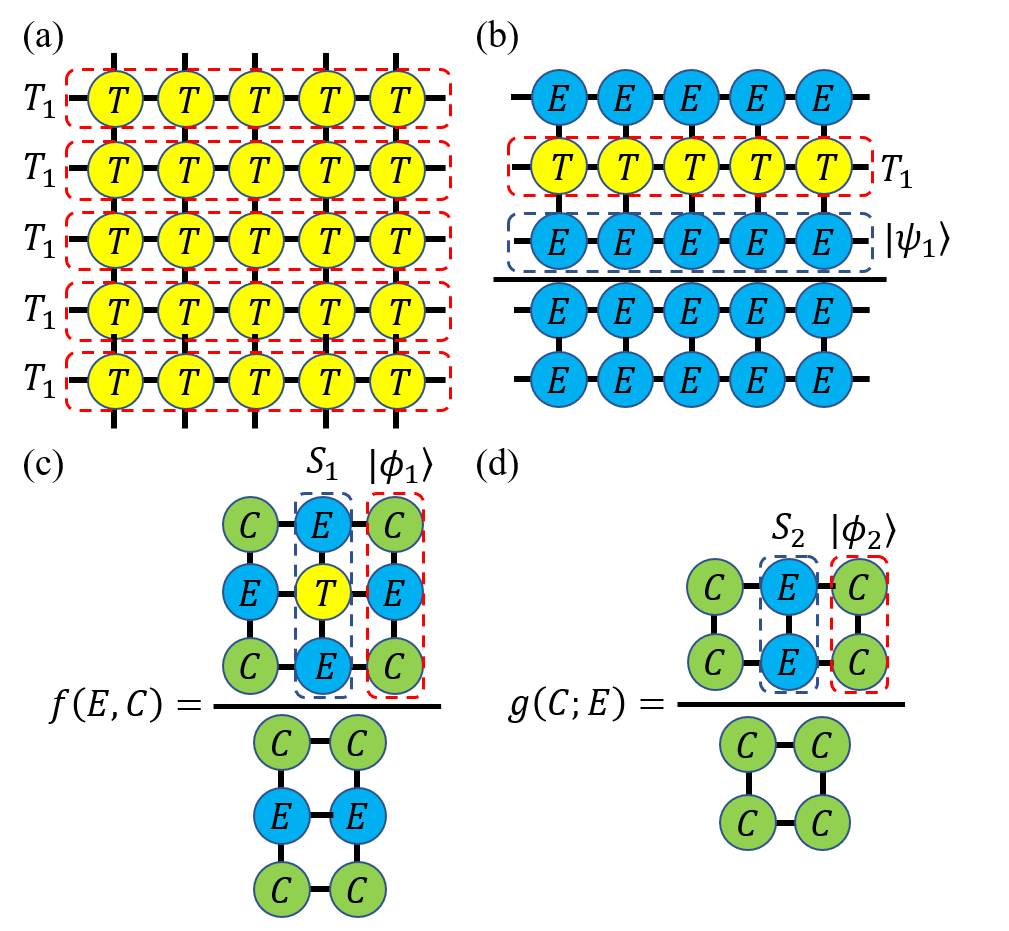}
		\caption{(Color online) Partition function represented by TNS: (a) infinite row tensors $T_1$; (b) represented by row tensor $T_1$ and its leading eigenvector $|\psi_1\rangle$; (c) numerator of (b), denoted by transfer matrix $S_1$ and its leading eigenvector $|\phi_1\rangle$; (d) denominator of (b), expressed by transfer matrix $S_2$ and its leading eigenvector $|\phi_2\rangle$.}
		\label{figure2}
	\end{figure}
	The infinite TNS is denoted by $\pounds$ with size $m\times n (m, n\to\infty)$, and also can be considered as $m$ row tensors $T_1$ as shown in Fig.\ \ref{figure2} (a). Then, the contraction of  $\pounds$ is reduced to the multiplication of $T_1$, and the partition function can be expressed as
	\begin{equation}\label{expressionZ1}
		\centering
		Z = {\text{Tr}}\prod_iT_{l_ir_iu_id_i} =  {\text{Tr}}(T_1)^m = \lambda^m,
	\end{equation}
	where $\lambda$ is the leading eigenvalue of row tensor $T_1$. Suppose $|\psi_1\rangle$ is the leading eigenvector of $T_1$, and represented as an uniform MPS composed of a local tensor $E$, as shown in Fig.\ \ref{figure2} (b), then Eq.\ (\ref{expressionZ1})  becomes
	\begin{equation}\label{expressionZ2}
		\centering
		Z = \left( \frac{\langle\psi_1|T_1|\psi_1\rangle}{\langle\psi_1|\psi_1\rangle} \right)^m.
	\end{equation}
	In this way, the problem is then {reduced} to contracting a triple-row infinite network. Further, one can construct a column transfer matrix $S_1$ and assume its leading eigenvector is $|\phi_1\rangle$, then the numerator of Eq.\ (\ref{expressionZ2}) is expressed as
	\begin{equation}\label{expressionf}
		\centering
		\langle\psi_1|T_1|\psi_1\rangle = \left(\frac{\langle\phi_1|S_1|\phi_1\rangle}{\langle\phi_1|\phi_1\rangle}\right)^n \equiv f^n,
	\end{equation}
	as shown in Fig.\ \ref{figure2} (c), where $|\phi_1\rangle$ has been represented as three-site MPS composed of local tensors $E$ and $C$ approximately. 
	Similarly, the denominator of Eq.\ (\ref{expressionZ2}) is factored into another transfer matrix $S_2$ and its leading eigenvector $|\phi_2\rangle$, see Fig.\ \ref{figure2} (d), as
	\begin{equation}\label{expressiong}
		\centering
		\langle\psi_1|\psi_1\rangle = \left(\frac{\langle\phi_2|S_2|\phi_2\rangle}{\langle\phi_2|\phi_2\rangle}\right)^n \equiv g^n.
	\end{equation}
	{Note} here we are assuming the system has both reflection and rotation symmetries, and the symmetric local tensor $C$ appears in both $|\phi_1\rangle$ and $|\phi_2\rangle$. 
	
	By introducing two quantities $f$ and $g$, as defined in Fig.~\ref{figure2} and in Eqs.~(\ref{expressionf}) and (\ref{expressiong}), the partition function can be expressed in the following,
	\begin{equation}\label{expressionZ3}
		\centering
		Z = \left(\frac{\langle\phi_1|S_1|\phi_1\rangle\langle\phi_2|\phi_2\rangle}{\langle\phi_1|\phi_1\rangle\langle\phi_2|S_2|\phi_2\rangle}\right)^{m\times n} = \left(\frac{f}{g}\right)^{mn}.
	\end{equation}
	\begin{figure}[htbp]
		\centering
		\includegraphics[width=0.48\textwidth]{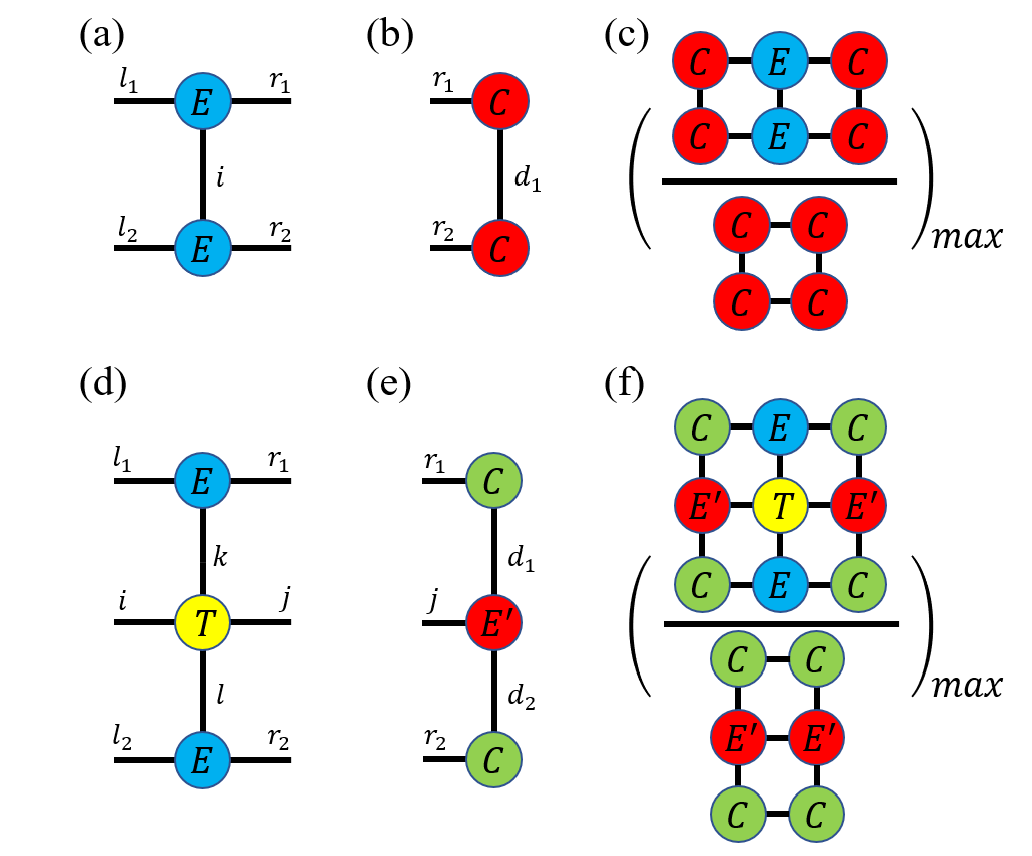}
		\caption{(Color online) (a) Transfer matrix $S_2$; (b) trial fixed-point vector $|\phi_2\rangle$; (c) maximize objective function to update corner matrix $C$; (d) transfer matrix $S_1$; (e) trial vector $|\phi_1\rangle$ constructed with updated $C$ and a variational tensor $E'$; (f) maximize objective function to optimize $E'$. Variational tensors are in red.}
		\label{figure3}
	\end{figure}

    {Essentially, the local tensors $E$ and $C$ correspond to the edge tensor and corner matrix exactly, as used to approximate the surrounding environment of the local tensor $T$ in the conventional CTMRG algorithm, where they're determined by some delicately designed iterations. Instead, according to Eq.~(\ref{expressiong}) and Eq.~(\ref{expressionZ3}), we propose to solve the following bilevel optimization problem
    \begin{eqnarray}
    	\max\limits_{E} \frac{f(E, C^*(E))}{g(C^*(E);E)}~ \text{s.t.} ~ C^*(E) = \mathop{\text{arg} \max}\limits_{C} g(C;E), \label{EqOpt}
    \end{eqnarray}
    whose solution $E$ and the corresponding $C$ construct the effective environment pursued in the conventional CTMRG iterations.} 

    This optimization problem can be solved efficiently by many methods, like the quasi-Newton method. However, without losing generality, we propose an iteration process in case that the derivative cannot be determined exactly. To be specific, in spirit of Baxter's original idea, we reformulate CTMRG in the following variational algorithm which aims to solve Eq.~(\ref{EqOpt}): (Hereinafter, $D$ and $\chi$ are used to denote the bond dimensions of local tensor $T$ and its environment, respectively.)
        
	\begin{enumerate}
		\item[(1)] Optimize $C$. Generate an arbitrary but symmetric rank-3 edge tensor $E_{\chi}$, construct the transfer matrix ${S_2}$, as shown in Fig.\ \ref{figure3} (a), and find the optimal $|\phi_2\rangle$, represented in Fig.\ \ref{figure3} (b), which maximizes Eq.\ (\ref{expressiong}) or Fig.\ \ref{figure3} (c).

		\item[(2)] Update $E'$. Construct transfer matrix $S_1$ with edge tensors $E$ and local tensor $T$ as in Fig.\ \ref{figure3} (d), from which its leading eigenvector $|\phi_1\rangle$ is obtained by optimizing the variational tensor $E'$ (see Fig.\ \ref{figure3} (e)) to maximize Eq.\ (\ref{expressionf}) or Fig.\ \ref{figure3} (f).
		
		\item[(3)] Replace $E$ in (1) by $E'$, and iterate the above two steps until the effective environment (i.e. $C$ and $E$) converges. 
	\end{enumerate}	

	Once $E$ and $C$ are obtained, the physical quantities, such as energy and magnetization etc., can be evaluated similarly as in the conventional CTMRG method. 
	
	\section{Results}
	
	\subsection{Dimer model}
	
	The dimer covering\cite{kasteleyn1963jmp}, also known as the perfect matching, {consists} of configurations satisfying the condition that all the vertices are the endpoints of one edge only. {It is exactly solvable and parameter free, but has extensive ground state degeneracy, thus provides an optimal platform to testify a new method}. As for a square lattice, the local tensor $T_{ijkl}$ of this model with bond dimension $D=2$ can be written as\cite{Vanderstraeten2018pre},
	\begin{equation}\label{dimerT}
		T_{ijkl}=\left\{
		\begin{array}{cl}
			1, & \text{only one index is 2;}\\
			0,  & \text{otherwise.} \\
		\end{array} \right.
	\end{equation}

\begin{figure}[htbp]
		\centering
		\includegraphics[width=0.3\textwidth]{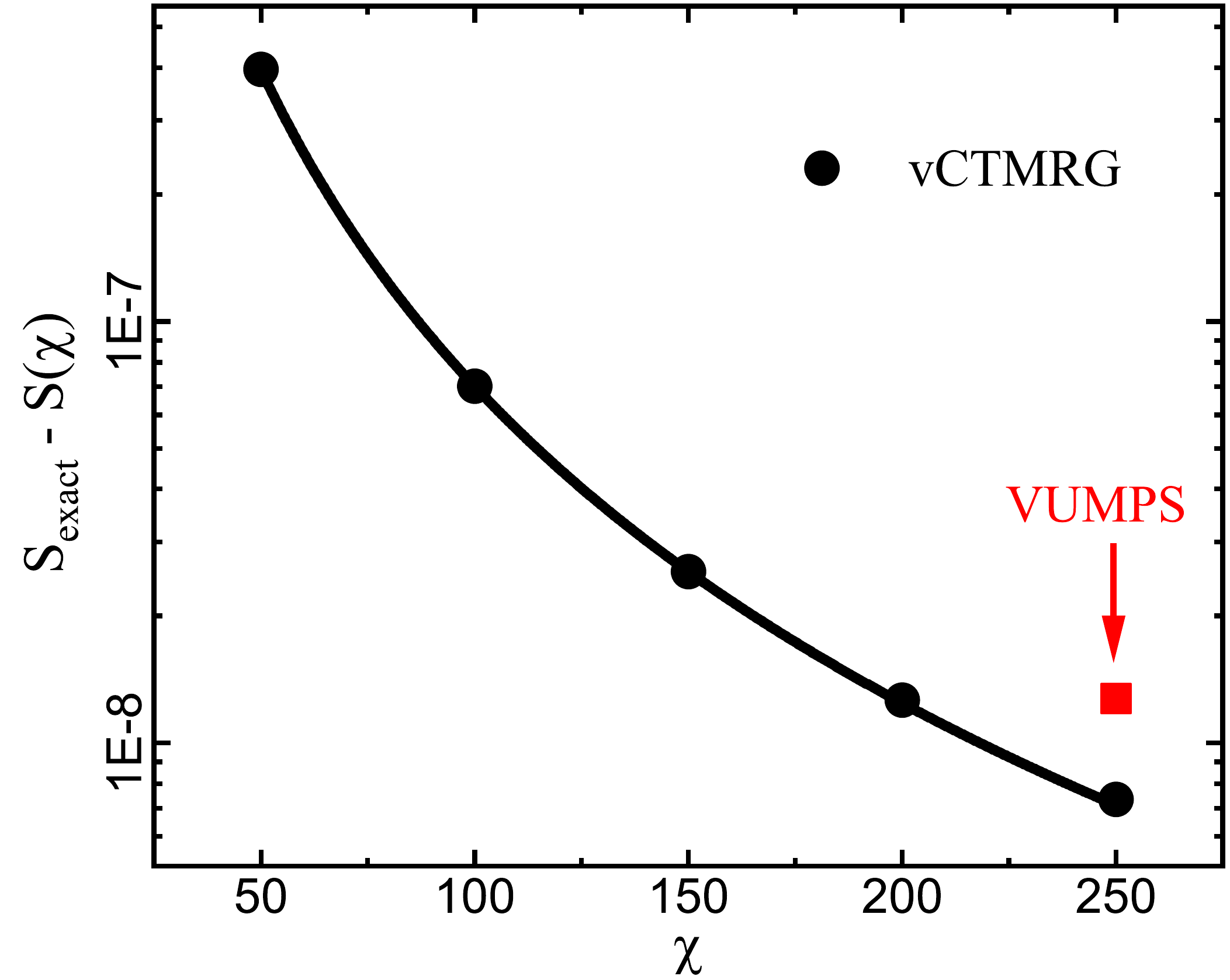}
		\caption{(Color online) Residual entropy $S(\chi)$ error of dimer model on the square lattice. The exact value $S_{\text{exact}}=G/\pi$, where $G\approx 0.916$ is the Catalan's constant \cite{kasteleyn1963jmp}.}
		\label{dimer}
	\end{figure}

	Following the above procedure described in Sect.\ \ref{SectMethod}, we calculate the partition function $Z(\chi)$  of this model for given bond dimension $\chi$, then obtain its residual entropy as $S(\chi)=\ln Z(\chi)$, which indicates the ground state degeneracy per site of the dimer configurations. The deviation from the theoretical value is shown in Fig.\ \ref{dimer}, with $\chi$ ranging from 50 to 250. The variational uniform matrix product state (VUMPS)\cite{Zauner-Stauber2018prb,Vanderstraeten2018pre} data with $\chi = 250$ is also included for comparison. Obviously, our variational estimation is \text{more accurate}. {More importantly, it shows clearly that the accuracy is systematically improved as increasing $\chi$, which verifies the variational nature of our method.}

	\subsection{Ising model}
	Ising model is the simplest statistical model which hosts phase transition. It considers only the interaction between nearest spins, with each spin taking $\pm1$. The local tensor $T$ is also of dimension $D=2$, and can be defiend as $T_{l_ir_iu_id_i} = \sum_{\alpha}W_{\alpha l_i}W_{\alpha r_i}W_{\alpha u_i}W_{\alpha d_i}$, where $W$ comes from the decomposition of Boltzmann factor, ($\beta$ is the inverse temperature)
	\begin{eqnarray}
		W=\begin{bmatrix} 
			\sqrt{\cosh\beta}, & \sqrt{\sinh\beta} \\
			 \sqrt{\cosh\beta}, & -\sqrt{\sinh\beta}
		 	\end{bmatrix}.
	\end{eqnarray}
		
	\begin{figure}[tbp]
		\centering
		\includegraphics[width=0.5\textwidth]{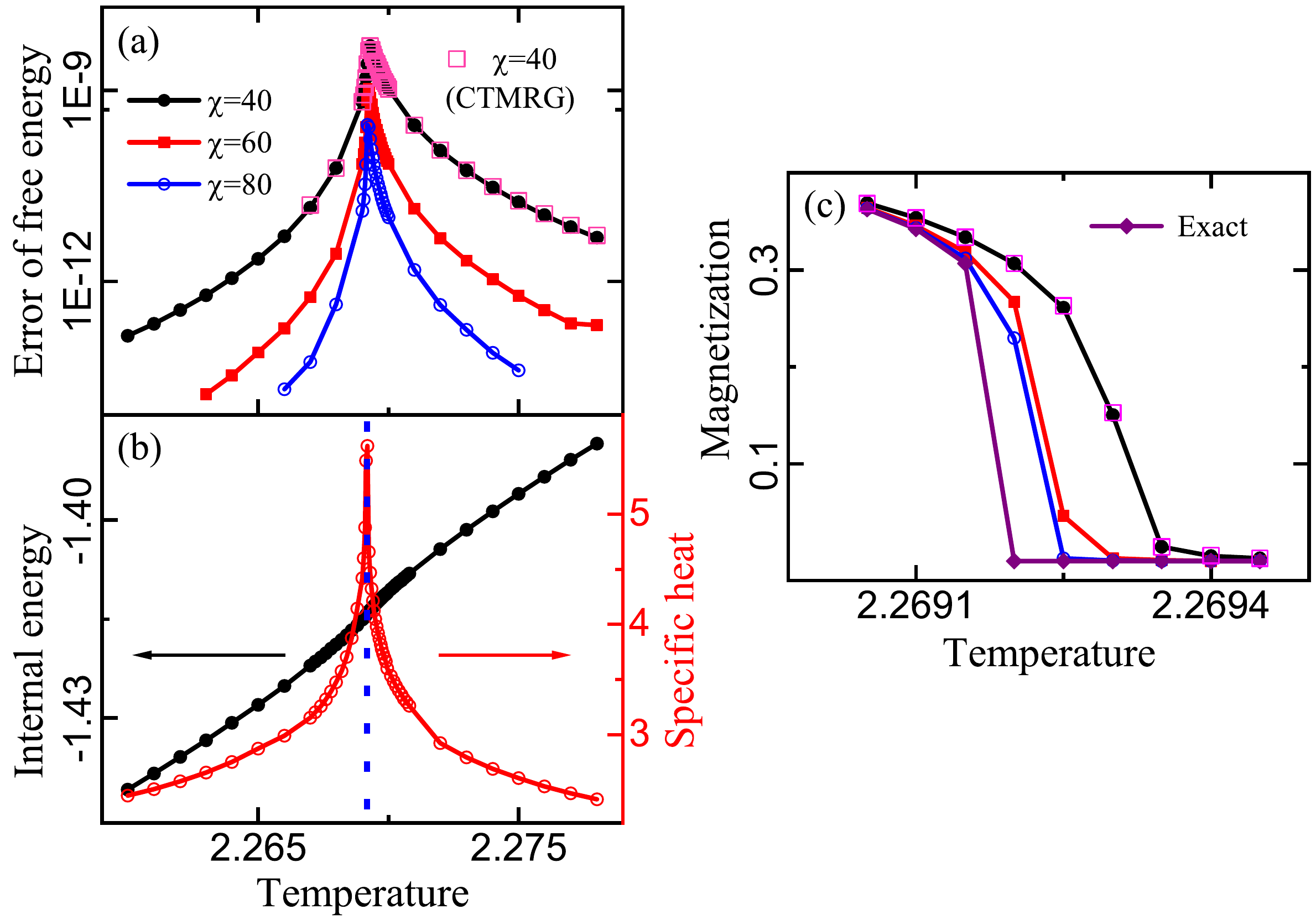}
		\caption{(Color online) (a) Absolute error of free energy obtained from vCTMRG method for different bond dimensions $\chi$. Data obtained from conventional CTMRG\cite{Orus2009prb}  is also included for comparison; (b) internal energy and specific heat for $\chi=80$, and the exact critical point is drawn as dotted vertical line at $T_c = 2/\ln(\sqrt{2}+1)$; (c) magnetization for different $\chi$, and compared to exact values and CTMRG data.}\label{ising}
	\end{figure}

	As shown in Fig.\ \ref{ising}, thermodynamic functions of this model on the square lattice are computed, where (a) is the absolute error of free energy for different bond dimensions $\chi$. It shows that for $\chi = 40$, the data coincides well with that obtained from conventional CTMRG method, as expected. In each curve, an abrupt peak apparently indicates the occurrence of a phase transition. Clearly, the error decreases monotonically with increasing $\chi$, and the critical point converges towards the theoretical value.
	This trend manifests itself in the magnetization curves in (c), where a moderate $\chi = 80$ result is almost identical to the exact one.	In Fig.\ \ref{ising} (b), the internal energy and the specific heat $C_V = {\partial U}/{\partial T}$ are given for $\chi=80$. As well-known, the phase transition in this model is of second order, and the investigated quantities above provide clear evidence and consistent estimations. The predicted critical point is $T_c = 2.26920(3)$, with error of order $10^{-5}$, and the accuracy can be further improved with larger dimension $\chi$.
	
	\subsection{Potts model}
	
	As an extension of Ising model, $Z(q)$-symmetric Potts model\cite{Potts1952mpcs, WuFY1982rmp} was proposed to consider each spin with $q(>2)$ choices in the plane, which thus has richer physics. The Hamiltonian is written as
	\begin{equation}\label{PottsT}
		H = -\sum_{\langle ij \rangle}\delta_{S_iS_j}.
	\end{equation}	
	\begin{figure}[htbp]
		\centering
		\includegraphics[width=0.48\textwidth]{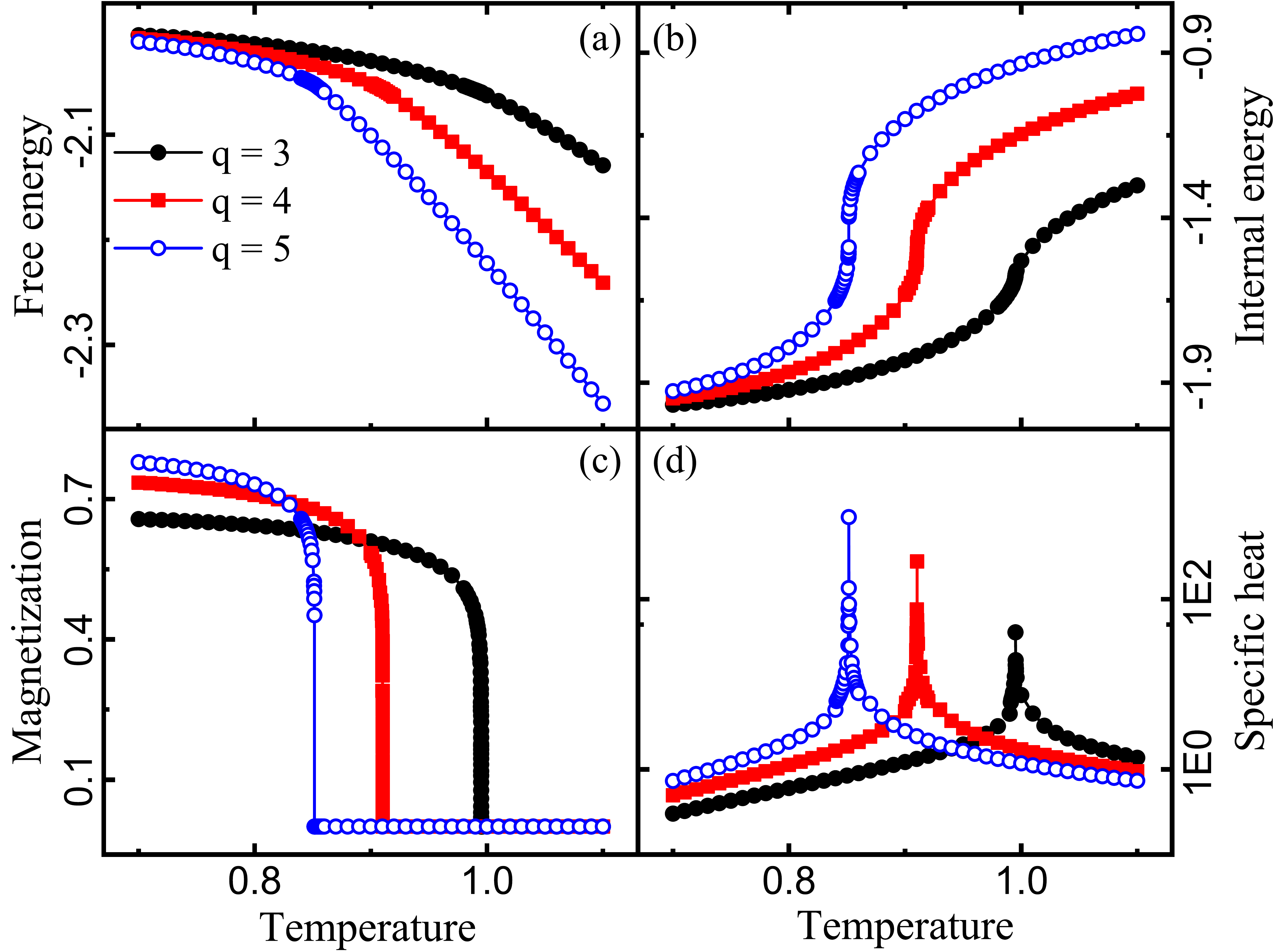}
		\caption{(Color online) Ferromagnetic Potts model on the square lattice for $q = 3, 4$ and $5$ cases with $\chi=20\times q$: (a) free energy; (b) internal energy; (c) magnetization; (d) specific heat.}
		\label{345potts}
	\end{figure}
	It is known that the model shows a second-order phase transition when $q<5$, and a first-order phase transition otherwise.
	
	Similarly, the free energy, internal energy, magnetization, and specific heat of $q = 3, 4, 5$ cases are calculated. As shown in Fig.\ \ref{345potts}, the clear transition can be observed from the magnetization or the specific heat, and the critical points are listed in Tab.\ \ref{tablepotts}, all consistent with the exact values\cite{WuFY1982rmp}. A slight discontinuity exhibits in the internal energy for $q=5$ case, different from $q=3, 4$ cases, and manifests itself in the magnetization and the specific heat, and the nature of different phase transitions is verified.
	
	\begin{table}[tb]
		\centering
		\caption{The critical points of $q=3,4,5$ ferromagnetic Potts model on the square lattice, comparing with exact values\cite{WuFY1982rmp}.}
		\label{tablepotts}
		\setlength{\tabcolsep}{2.1mm}{
			\begin{tabular}{cccc}
				\hline \hline & Calculated $T_c$ & Exact & Relative error \\
				\hline $q=3$ & 0.99505(5) & $1/\ln(\sqrt{3}+1)$ & $7.7529\times 10^{-5}$ \\
				\hline $q=4$ & 0.91025(5) & $1/\ln(3)$ & $1.1836\times 10^{-5}$ \\
				\hline $q=5$ & 0.85155(5) & $1/\ln(\sqrt{5}+1)$ & $2.5411\times 10^{-5}$ \\
				\hline \hline
		\end{tabular}}
	\end{table}
	Once the critical temperature is obtained, one can calculate the critical exponents $\alpha$ and $\gamma$ of the second-order phase transition. They determine the singular behavior of specific heat and  magnetization near the critical point, i.e., $C_V \sim t^{-\alpha}$ and $m \sim t^{\gamma}$ respectively, where $t=|{(T_c-T)}/{T_c}|$ is the reduced temperature. The results are shown in Tab.\ \ref{tablecriticalpotts}, both in consonance with exact values. The high precision estimation of both critical points and critical exponents of this model further demonstrates the effectiveness and efficiency of our proposed algorithm.
	
	\begin{table}[tb]
		\centering
		\caption{The critical exponents $\alpha$ and $\gamma$ of ferromagnetic Potts model on the square lattice, with the exact values\cite{WuFY1982rmp} included in parentheses.}
		\label{tablecriticalpotts}
		\setlength{\tabcolsep}{2.1mm}{
			\begin{tabular}{ccc}
				\hline \hline & $\alpha$ & $\gamma$  \\
				\hline $q=3$ & 0.3314($\frac{1}{3}$) & 0.1118($\frac{1}{9}$)  \\
				\hline $q=4$ & 0.6359($\frac{2}{3}$) & 0.0889($\frac{1}{12}$) \\
				\hline \hline
		\end{tabular}}
	\end{table}
	\subsection{Clock model}
	
	An even more interesting and complex model is the $q$-state clock model, i.e. the discrete version of XY model\cite{YuJF2014pre}, because of KT transition\cite{kosterlitz1973jpc, kosterlitz1974jpc, Jose1977prb} with exotic topological excitation there. Many believe that when $2<q\leq4$, the only transition is of second-order Landau type; {while} if $q>4$, there are two transitions, both of KT type\cite{kosterlitz1973jpc, kosterlitz1974jpc, Jose1977prb} with quasi-long-range-ordered critical phase in between the low temperature ordered-phase and high temperature paramagnetic-phase. Here, we focus on $q = 3,4,5$ cases, covering both types of phase transitions. 
	The Hamiltonian reads
	\begin{equation}\label{clockT}
		\centering
		H = -\sum_{\langle ij \rangle}\cos(\theta_i-\theta_j),
	\end{equation}
	where the summation runs over all the nearest neighbors, and $\theta_i = \frac{2\pi k}{q}$ is the spin angle at site $i$, and integer $k$ ranges from $0$ to $q-1$.

	\begin{figure}[htbp]
		\centering
		\includegraphics[width=0.45\textwidth]{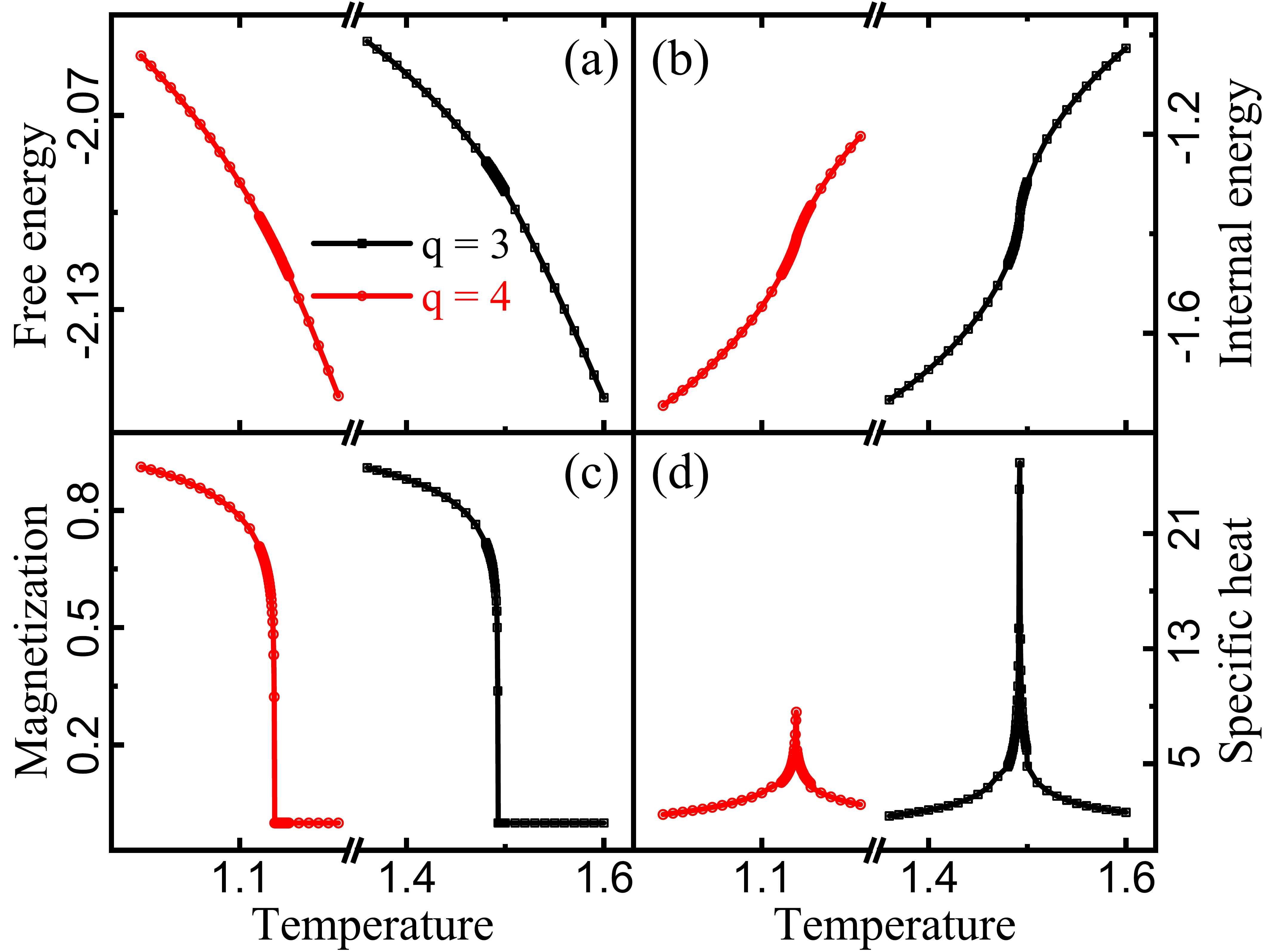}
		\caption{(Color online) $q=3,4$-state clock model on the square lattice with $\chi=20\times q$: (a) free energy; (b) internal energy; (c) magnetization; (d) specific heat.}
		\label{34clock}
	\end{figure}
	
	The results of $q =3$ and $4$ cases are shown in Fig.\ \ref{34clock}. Again the second-order phase transition can be seen from the divergence of magnetization\cite{Tomita2002prb, Rastelli2004prb, Borisenko2011pre} $m=\sqrt{\langle \cos\theta\rangle^2+\langle \sin\theta\rangle^2}$ and specific heat. The corresponding transition temperatures are listed in Tab.\ \ref{tableclock}, both consistent with the exact values\cite{Kramers1941pr}. Further, the critical exponents are computed and given in Tab.\ \ref{tablecriticalclock}, all agreeing well with the theoretical values.
	
	\begin{table}[tb]
		\centering
		\caption{The critical points of $q=3,4$ clock model on the square lattice, compared to exact values\cite{Kramers1941pr}.}
		\label{tableclock}
		\setlength{\tabcolsep}{2.1mm}{
			\begin{tabular}{cccc}
				\hline \hline & Calculated $T_c$ & Exact & Relative error \\
				\hline $q=3$ & 1.4925(5) & $3/[2\ln(\sqrt{3}+1)]$ & $2.7276\times 10^{-5}$ \\
				\hline $q=4$ & 1.1350(5) & $1/\ln(\sqrt{2}+1)$ & $3.5902\times 10^{-4}$ \\
				\hline \hline
		\end{tabular}}
	\end{table}
	
	\begin{table}[htb]
		\centering
		\caption{The critical exponents $\alpha$ and $\gamma$ of ferromagnetic clock model for $q=3, 4$ on the square lattice, with exact values\cite{WuFY1982rmp, Suzuki1967ptp} shown in parentheses.}
		\label{tablecriticalclock}
		\setlength{\tabcolsep}{2.1mm}{
			\begin{tabular}{ccc}
				\hline \hline & $\alpha$ & $\gamma$  \\
				\hline $q=3$ & 0.3696($\frac{1}{3}$) & 0.1111($\frac{1}{9}$)  \\
				\hline $q=4$ & 0(0) & 0.1263($\frac{1}{8}$) \\
				\hline \hline
		\end{tabular}} 
	\end{table}	
	
\begin{figure}[htb]
		\centering
		\includegraphics[width=0.5\textwidth]{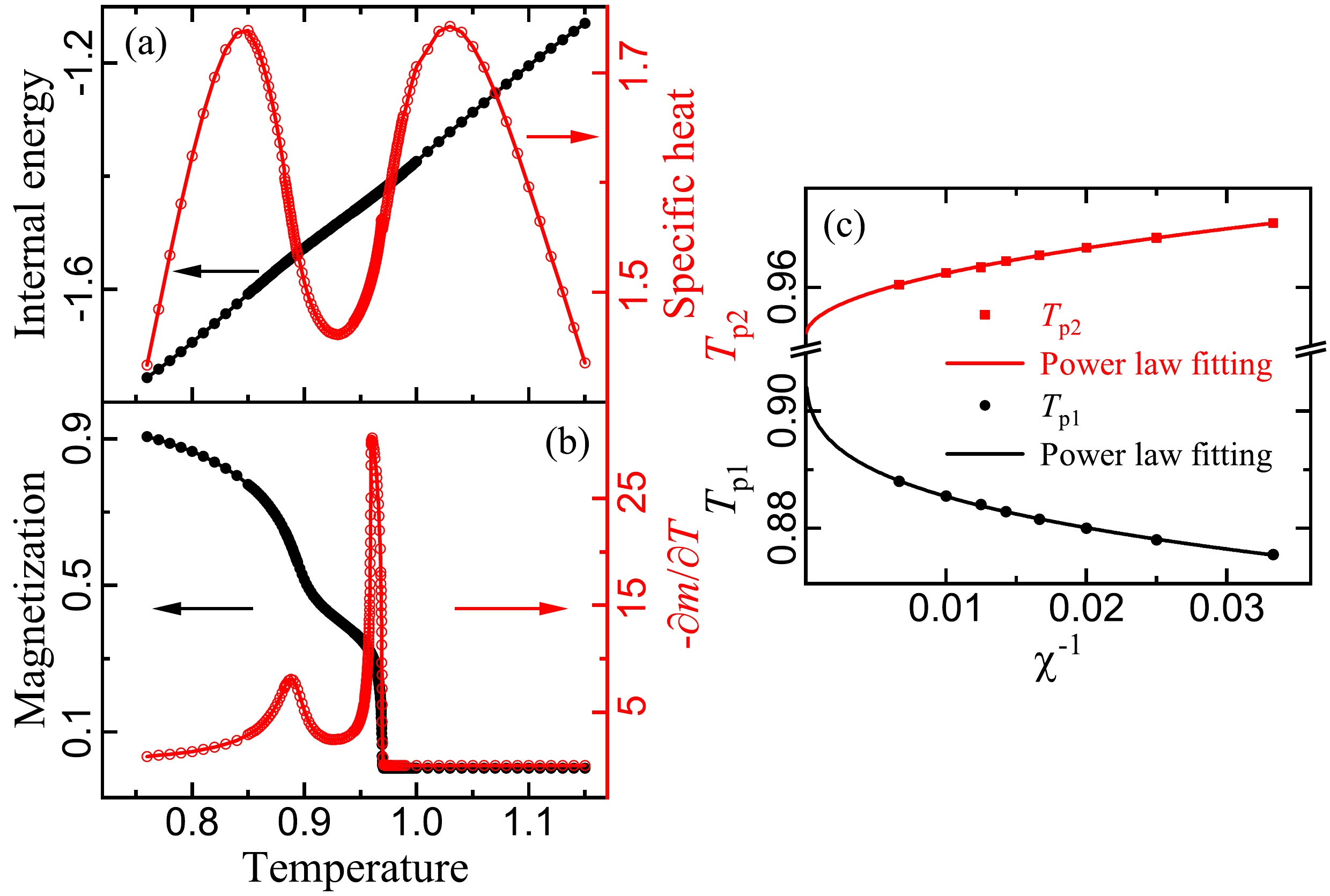}
		\caption{(Color online) $5$-state clock model on the square lattice with dimension $\chi=150$: (a) internal energy and specific heat; (b) magnetization and $-\partial m/\partial T$; (c) peak positions of $-\partial m/\partial T$ versus $\chi^{-1}$, and power-law fittings are performed to extrapolate the critical temperatures at $T_{c1} = 0.9099(45)$ and $T_{c2} = 0.9516(11)$, respectively.}\label{q5clock}
	\end{figure}

	While, the $q=5$ case is totally different, as can be observed from Fig.\ \ref{q5clock} (a) specific heat or (b) magnetization, with dimension $\chi=150$. The quantity is continuous without any singularity. There are two transitions, as discussed in Ref.\ \onlinecite{ChenY2020prb}, and the upper one can be {more easily} captured by the magnetic susceptibility than the lower one. Following the idea in 
Ref.\ \onlinecite{ChenY2020prb}, the function $\partial m /\partial T$ is computed, where two clear peaks show up and signal two phase transitions. We further vary the dimension $\chi$ to obtain the corresponding peak temperature, as shown in Fig.\ \ref{q5clock} (c), and perform power-law fittings to obtain the transition temperatures as $T_{c1}=0.9099(45)$ and $T_{c2}= 0.9516(11)$, both consistent with other estimations as listed in Tab.\ \ref{tableq5}.
	\begin{table}[htb]
		\centering
		\caption{Comparison of the critical temperatures $T_{c1}$ and $T_{c2}$ by different
			methods for the $5$-state clock model.}
		\label{tableq5}
		\setlength{\tabcolsep}{2.1mm}{
			\begin{tabular}{ccc}
				\hline \hline & $T_{c1}$ & $T_{c2}$  \\
				\hline MC\cite{Borisenko2011pre} & 0.90514(9) & 0.95147(9) \\
				DMRG\cite{Chatelain2014jsm} & 0.914(12) & 0.945(17)  \\
				HOTRG\cite{ChenY2018cpb} & 0.9029(1) & 0.9520(1) \\
				VUMPS\cite{LiZQ2020pre} & 0.9059(2) & 0.9521(2) \\
				\hline vCTMRG & 0.9099(45) & 0.9516(11) \\
				\hline \hline
		\end{tabular}} 
	\end{table}	

	\section{Discussions and Conclusions}
	
	In summary, we propose a variational CTMRG algorithm which amounts to solving a bilevel optimization problem, whose solution corresponds to the fixed-point environment in the conventional CTMRG iteration process.  The algorithm has been testified in a series of classical models, including an exactly solvable model with extensive ground state degeneracy, and also systems hosting first-order, second-order, as well as KT transitions. The obtained results, such as critical points and critical exponents, agree very well with either the exact results or the previous studies by other methods. 
	
    The method realizes the variational idea in terms of CTM tensors in the context of tensor networks for the first time, and the biggest difference from the conventional CTMRG is the reformulation of CTMRG as an optimization problem.	For example, the conventional CTMRG method either performs the essential power iterations in two directions alternatively \cite{Orus2009prb,Corboz2014prl},  or solves the coupled fixed-point equations in two directions through some canonical transformations\cite{Fishman2018prb}, while neither directly touches the partition function itself. In other words, there is no such kind of guidance like Eq.~(\ref{EqOpt}) in the iteration process. In this sense, the new formulation proposed here is variational, and more efficient.  
    
    In this work, the local tensor has both reflection and rotational symmetries as in most statistical models, but the extension to systems without these symmetries can be expected. In those cases, the left leading eigenvector and the right leading eigenvector can be different, thus four different edges and corners should be involved, and the corresponding optimization conditions should be also added to the target function. This is especially useful when considering a quantum lattice model, where the two symmetries can be broken, like a multi-sublattice system. It can be expected that the variational nature of the proposed method can improve the convergence and reliability of tensor network methods.
    
    The last but not the least important, a significant advantage of this method is that it can be extended straightforwardly to three-dimensional lattice. In that case, the optimization problem becomes trilevel instead of bilevel, since we need further a face tensor besides the corner and edge tensors. Roughly speaking, the optimization process is quite similar as in two-dimension, but the target function is more complicated, and the computational cost will be much higher. In view of the fact that the conventional CTMRG iteration cannot be easily applied to three-dimensional lattice \cite{Orus2012prb}, the variational CTMRG method is a promising approach for the study of three-dimensional systems, which we leave as a future pursuit.
    
	\section{Acknowledgments}
	
	 We thank Prof. Tao Xiang for helpful discussions and comments. The work is supported by National R\&D Program of China (Grants No. 2017YFA0302900), the National Natural Science Foundation of China (Grants No. 11774420 and No. 12134020), the Natural Science Foundation of Hunan Province (No. 851204035), and the Fundamental Research Funds for the Central Universities and the Research Funds of Renmin University of China (Grants No. 20XNLG19). X.F.L and Y.F.F contribute equally to this {work}.

	\bibliography{Bibtex}

\end{document}